\documentclass[12pt]{article}
\usepackage{amsfonts}
\usepackage{amssymb}

\def\hybrid{\topmargin -20pt    \oddsidemargin 0pt
        \headheight 0pt \headsep 0pt
        \textwidth 6.25in       
        \textheight 9.25in       
        \marginparwidth .875in
        \parskip 5pt plus 1pt   \jot = 1.5ex}

\hybrid

\def\baselinestretch{1.2}

\catcode`\@=11

\def\marginnote#1{}
\newcount\hour
\newcount\minute
\newtoks\amorpm
\hour=\time\divide\hour by60
\minute=\time{\multiply\hour by60 \global\advance\minute by-\hour}
\edef\standardtime{{\ifnum\hour<12 \global\amorpm={am}%
        \else\global\amorpm={pm}\advance\hour by-12 \fi
        \ifnum\hour=0 \hour=12 \fi
        \number\hour:\ifnum\minute<10 0\fi\number\minute\the\amorpm}}
\edef\militarytime{\number\hour:\ifnum\minute<10 0\fi\number\minute}

\def\draftlabel#1{{\@bsphack\if@filesw {\let\thepage\relax
   \xdef\@gtempa{\write\@auxout{\string
      \newlabel{#1}{{\@currentlabel}{\thepage}}}}}\@gtempa
   \if@nobreak \ifvmode\nobreak\fi\fi\fi\@esphack}
        \gdef\@eqnlabel{#1}}
\def\@eqnlabel{}
\def\@vacuum{}
\def\draftmarginnote#1{\marginpar{\raggedright\scriptsize\tt#1}}

\def\draft{\oddsidemargin -.5truein
        \def\@oddfoot{\sl preliminary draft \hfil
        \rm\thepage\hfil\sl\today\quad\militarytime}
        \let\@evenfoot\@oddfoot \overfullrule 3pt
        \let\label=\draftlabel
        \let\marginnote=\draftmarginnote
   \def\@eqnnum{(\theequation)\rlap{\kern\marginparsep\tt\@eqnlabel}%
\global\let\@eqnlabel\@vacuum}  }

\def\preprint{\twocolumn\sloppy\flushbottom\parindent 2em
        \leftmargini 2em\leftmarginv .5em\leftmarginvi .5em
        \oddsidemargin -.5in    \evensidemargin -.5in
        \columnsep .4in \footheight 0pt
        \textwidth 10.in        \topmargin  -.4in
        \headheight 12pt \topskip .4in
        \textheight 6.9in \footskip 0pt
        \def\@oddhead{\thepage\hfil\addtocounter{page}{1}\thepage}
        \let\@evenhead\@oddhead \def\@oddfoot{} \def\@evenfoot{} }

\def\numberbysection{\@addtoreset{equation}{section}
        \def\theequation{\thesection.\arabic{equation}}}

\def\underline#1{\relax\ifmmode\@@underline#1\else
        $\@@underline{\hbox{#1}}$\relax\fi}

\def\titlepage{\@restonecolfalse\if@twocolumn\@restonecoltrue\onecolumn
     \else \newpage \fi \thispagestyle{empty}\c@page\z@
        \def\thefootnote{\fnsymbol{footnote}} }

\def\endtitlepage{\if@restonecol\twocolumn \else \newpage \fi
        \def\thefootnote{\arabic{footnote}}
        \setcounter{footnote}{0}}  

\catcode`@=12
\relax

\def\figcap{\section*{Figure Captions\markboth
        {FIGURECAPTIONS}{FIGURECAPTIONS}}\list
        {Figure \arabic{enumi}:\hfill}{\settowidth\labelwidth{Figure
999:}
        \leftmargin\labelwidth
        \advance\leftmargin\labelsep\usecounter{enumi}}}
 \relax
\def\tablecap{\section*{Table Captions\markboth
        {TABLECAPTIONS}{TABLECAPTIONS}}\list
        {Table \arabic{enumi}:\hfill}{\settowidth\labelwidth{Table
999:}
        \leftmargin\labelwidth
        \advance\leftmargin\labelsep\usecounter{enumi}}}
 \relax
\def\reflist{\section*{References\markboth
        {REFLIST}{REFLIST}}\list
        {[\arabic{enumi}]\hfill}{\settowidth\labelwidth{[999]}
        \leftmargin\labelwidth
        \advance\leftmargin\labelsep\usecounter{enumi}}}
 \relax
\makeatletter
\newcounter{pubctr}
\def\publist{\@ifnextchar[{\@publist}{\@@publist}}
\def\@publist[#1]{\list
        {[\arabic{pubctr}]\hfill}{\settowidth\labelwidth{[999]}
        \leftmargin\labelwidth
        \advance\leftmargin\labelsep
        \@nmbrlisttrue\def\@listctr{pubctr}
        \setcounter{pubctr}{#1}\addtocounter{pubctr}{-1}}}
\def\@@publist{\list
        {[\arabic{pubctr}]\hfill}{\settowidth\labelwidth{[999]}
        \leftmargin\labelwidth
        \advance\leftmargin\labelsep
        \@nmbrlisttrue\def\@listctr{pubctr}}}
 \relax
\makeatother
\newskip\humongous \humongous=0pt plus 1000pt minus 1000pt

\newif\ifdtup

\relax

\def\be{\begin{equation}}
\def\ee{\end{equation}}
\def\ba{\begin{eqnarray}}
\def\ea{\end{eqnarray}}

\def\del{\partial}

\def\r{\rho}
\def\a{\alpha}

\def\G{\Gamma}
\def\d{\delta}
\def\D{\Delta}
\def\e{\epsilon}

\def\P{\Pi}

\def\th{\theta}
\def\Th{\Theta}

\def\om{\omega}

\def\s{\sigma}
\def\S{\Sigma}

\def\cL{{\cal L}}

\def\no{\noindent}

\def\qq{\qquad}

\def\IR{\relax{\rm I\kern-.18em R}}

\def \ha {{1\over 2}}

\def \ov {\over}

\def\II{\relax{\rm 1\kern-.35em1}}
\def\IR{\relax{\rm I\kern-.18em R}}
\def\inv{^{\raise.15ex\hbox{${\scriptscriptstyle -}$}\kern-.05em 1}}

\def\cL{{\cal L}}

\begin{document}
\renewcommand{\theequation}{\arabic{equation}}
\renewcommand{\theequation}{\thesection.\arabic{equation}}

\newcommand{\beq}{\begin{equation}}
\newcommand{\eeq}[1]{\label{#1}\end{equation}}
\newcommand{\ber}{\begin{eqnarray}}
\newcommand{\eer}[1]{\label{#1}\end{eqnarray}}
\newcommand{\eqn}[1]{(\ref{#1})}
\begin{titlepage}
\begin{center}

\hfill CPHT-RR070.1105 \vskip -.3 cm
\hfill hep--th/0512251\\

\vskip  0.5in

{\large \bf NS5-branes on an ellipsis and novel \break marginal
deformations with parafermions${}^\ast\!$}

\vskip 0.4in

{\bf P. Marios Petropoulos$^1$}\phantom{x} and\phantom{x} {\bf
Konstadinos Sfetsos}$^{2,1}$ \vskip 0.1in

${}^1\!$ Centre de Physique Th{\'e}orique, Ecole Polytechnique${}^\dagger\!$\\
91128 Palaiseau Cedex, France\\
{\footnotesize{\tt marios@cpht.polytechnique.fr}}

\vskip .2in

${}^2\!$
Department of Engineering Sciences, University of Patras\\
26110 Patras, Greece\\
{\footnotesize{\tt sfetsos@des.upatras.gr}}\\

\end{center}

\vskip .3in

\centerline{\bf Abstract}

\no We consider NS5-branes distributed along the circumference of
an ellipsis and explicitly construct the corresponding
gravitational background. This provides a continuous line of
deformations between the limiting cases, considered before, in
which the ellipsis degenerates into a circle or into a bar. We
show that a slight deformation of the background corresponding to
a circle distribution into an ellipsoidal one is described by a
novel non-factorizable marginal perturbation of bilinears of
dressed parafermions. The latter are naturally defined for the
circle case since, as it was shown in the past, the
background corresponds to an orbifold of the exact conformal field
theory coset model $SU(2)/U(1) \times SL(2,\mathbb{R})/U(1)$. We
explore the possibility to define parafermionic objects at generic
points of the ellipsoidal families of backgrounds away from the
circle point. We also discuss a new limiting case in which the
ellipsis degenerates into two infinitely stretched parallel bars
and show that the background is related to the Eguchi--Hanson
metric, via T-duality.

\vfil
\hrule width 6.7cm \vskip.1mm{\small \small \small \noindent
$^\ast$\ Research partially supported by the EU under the
contracts MEXT-CT-2003-509661,
MRTN-CT-2004-005104 and MRTN-CT-2004-503369.\\
$^\dagger$\  Unit{\'e} mixte  du CNRS et de  l'Ecole
Polytechnique, UMR 7644.}
\end{titlepage}
\vfill
\newpage
\setcounter{footnote}{0}
\renewcommand{\thefootnote}{\arabic{footnote}}


\setcounter{section}{0}

\def\baselinestretch{1.2}
\baselineskip 17.5 pt
\noindent


\tableofcontents

\section{Introduction}

The investigation of backgrounds created by brane configurations
has allowed progress in various directions of string theory. In
most situations, however, this progress is limited by the
supergravity approximation. Even in the absence
of Ramond--Ramond fluxes, exact string solutions with a clear
brane interpretation are not numerous
\cite{Callan}--\cite{Sfetsos:1999pq}. An interesting and
unexpected realization of exact string background was provided in
\cite{sfet1}, by distributing $N$ NS5-branes on a circle, either on a
discrete set of points or continuously. In the latter case the
geometry, antisymmetric tensor and dilaton backgrounds turn out to
be T-dual to those of the $SL(2,\mathbb{R})/U(1) \times
SU(2)/U(1)$ conformal sigma model.

An exact worldsheet conformal sigma model is in general the seed
for generating a wider continuous or discrete moduli space of
exact string solutions. This includes in particular continuous
deformations induced by marginal worldsheet operators.\footnote{The
issue of marginal deformations in conformal models has been
discussed prolifically in the literature. We cannot be exhaustive
here.} Examples of this
kind and in the present framework were carried out exhaustively \cite{Forste:1994wp}--\cite{Israel}, as
e.g. the $SU(2)/U(1) \times SL(2,\mathbb{R})/U(1)$ model, which
is known to be connected to the $SU(2)\times
SL(2,\mathbb{R})$ by an exact marginal deformation triggered by
bilinears in the $SU(2)$ and $SL(2,\mathbb{R})$ currents.
It turns out though, that along the lines of deformation one
usually looses track of the brane-sources, except in a few
examples such as null deformations of $SL(2,\mathbb{R})$ or
related models. We know for example that
a T-dual version of $SU(2)/U(1) \times SL(2,\mathbb{R})/U(1)$
is also related to $SU(2)\times
SL(2,\mathbb{R})$ by a marginal deformation
generated by a bilinear of null combinations of  $SU(2)$ and
$SL(2,\mathbb{R})$ currents. It connects the background of a
continuous distribution of NS5-branes on a circle and that of
NS5-branes at a point with
F1 spread in their world-volume \cite{Israel:2003ry, Israel}.

The original purpose of the present work is to present an example where
an exact background is deformed by continuously deforming the distribution
of the brane-sources, allowing thereby to be all the way in contact with
the physical interpretation of the background fields. To be specific,
we will consider NS5-branes distributed over a circle which is deformed
into an ellipsis. Such a deformation is fully controllable at the
supergravity level and embraces known geometries such as the
Eguchi--Hanson solution \cite{EH}. The remarkable outcome of our analysis
is that the departure from the circle is also driven by a marginal
operator of the underlying $SL(2,\mathbb{R})/U(1) \times
SU(2)/U(1)$ conformal field theory. Such a result is unexpected
because of the absence of currents in these coset models. Despite that,
the compact parafermions of $SU(2)/U(1)$ can be appropriately dressed by
the non-compact fields of $SL(2,\mathbb{R})/U(1)$ and deliver a novel kind of
operator with anomalous dimension two, which is \emph{not factorizable}
in terms of holomorphic and anti-holomorphic currents. Per se this is a noticeable
observation because no example of this type seems to appear elsewhere.

The supergravity solution we are describing exists away from the vicinity
of the circular distribution, for any finite value of the eccentricity
parameter of the ellipsis. This suggests the marginal operator be exact
and always present in the worldsheet theory of the ``elliptic'' background.

We suggest \emph{generalized compact and non-compact parafermions},
extensions of the ordinary $SU(2)$ and $SL(2,\mathbb{R})$ ones, and we show that
the generic deformation is driven by dressed bilinears of these compact
and non-compact generalized parafermions. The existence of this operator at
any point of the elliptic deformation indicates that it is
\emph{exactly marginal} and that the corresponding worldsheet theory is exactly
conformal. A rigorous proof of this statement requires the computation of
the exact anomalous dimension of this \emph{non-factorizable} operator,
which is left for future work.

\section{Supergravity solution for NS5-branes on an ellipsis}

\setcounter{equation}{0}
\renewcommand{\theequation}{\thesection.\arabic{equation}}
\renewcommand{\theequation}{\arabic{equation}}

\subsection{A constructive approach}

The general form of the metric representing the gravitational
field of a large number $N$ of NS5-brane gravitational solution is
\be
\mathrm{d}s^2= \mathrm{d}s^2\left(E^{(1,5)}\right) + H \mathrm{d}s^2\left(\mathbb{R}^4\right)\ .
\label{jw11}
\ee
This is
promoted to a solution of the field equations when it is
supplemented by a closed three-form given by
\be
H_{ijk}=\e_{ijk}{}^l\del_l H\ ,
\label{jw1}
\ee
where the indices
are lowered and raised with the flat metric in $\mathbb{R}^4$, and
a dilaton field
\be
\mathrm{e}^{2\Phi}=H\ .
\label{dilr}
\ee
Satisfying the field
equations and preserving half of maximum supersymmetry requires
that the function $H$ is harmonic in $\mathbb{R}^4$ and in general
be given by
\be
H({\bf x})=N \int_{\mathbb{R}^4} \mathrm{d}^4{x'}
{\r({\bf x})\ov |{\bf x}-{\bf x'}|^2}\ ,
\label{harrm}
\ee
for some everywhere
positive density function $\r({\bf x})$ normalized to 1. Since we are solely
interested in the field theory--near horizon limit we have
omitted the unity from the above expression for the harmonic function.
If all centers coincide then the harmonic function above is simply
$H=N/r^2$, where $r$ is the radial variable in $\mathbb{R}^4$. In
this case the non-trivial four-dimensional part of the background
is described by the exact conformal field theory (CFT)
given as the direct product of the
$SU(2)_N$ Wess--Zumino--Witten (WZW) model with a free boson
with a background charge
\cite{Callan}. In the generic case in which the centers are randomly
distributed the symmetry of $\mathbb{R}^4$ is completely broken
and finding a CFT associated to the background seems hopeless.
In \cite{sfet1} the background corresponding to NS5-branes uniformly
distributed on a canonical $N$-polygon, was explicitly
constructed. In this case a $U(1)\times Z_N$ symmetry is preserved
where the $U(1)$
factor corresponds to rotations in the two-dimensional space transverse to
the plane where the NS5-branes are distributed. On the other hand
the $Z_N$ factor is due to the discreteness of
the distribution. In the continuum limit this factor becomes
another $U(1)$ and the NS5-branes are then continuously and
uniformly distributed on the circumference of a circle.
Moreover, using a T-duality, it was shown, that there
is an exact CFT corresponding to it given by an orbifold of the
$SU(2)/U(1)\times SL(2,\mathbb{R})/U(1)$ coset model \cite{sfet1}
(see also \cite{Giveon} and for further details \cite{KKPRproc, Israel}).

In this paper we will consider NS5-branes
distributed on the circumference of an ellipsis with axes $\ell_1$
and $\ell_2$ and density distribution
\be
\r({\bf x})= {1\ov \pi \ell_1 \ell_2}\
\d\left(1-{x_3^2\ov\ell_1^2}-{x_4^2\ov\ell_2^2}\right)\d(x_1)\d(x_2)
\ ,
\label{dits}
\ee
where we confine our discussion to the continuum limit.
The symmetry of the background is inherited
by that of the brane distribution. Therefore we expect that the
background will have a reduced $U(1)\times Z_2$ symmetry, where
the first factor is the same as in the case of the circle
distribution, namely it will correspond to rotations of the transverse to
the NS5-branes plane.
The $Z_2$ factor essentially interchanges the two axes of the elllipsis.
We will be able to describe the deformation of a circle into
an ellipsis as a novel non-factorizable perturbation by compact parafermion bilinears,
appropriately dressed so that the perturbation is marginal.
Moreover, their group-theoretical transformation properties under the broken $U(1)$
symmetry factor will explain, as we shall see, the reduced
symmetry from a CFT view point.

To proceed, we follow \cite{BakSfe,BBS} and parameterize in a quite general
way the coordinates $x_i$ it terms of four units vectors $\hat x_i$'s
restricted on the unit $S^3$, i.e. $\sum_{i=1}^4 \hat x_i^2 =1$,
as
\be {\bf x} = \left(\sqrt{r^2-b_1}\ \hat x_1,\sqrt{r^2-b_2}\ \hat
x_2, \sqrt{r^2-b_3}\ \hat x_3, \sqrt{r^2-b_4}\ \hat x_4\right)\ .
\ee
The constants $b_i$ can be taken without loss of generality to be
such that $b_1\geq b_2\geq b_3\geq b_4$. This parameterization
dictates that the radial variable $r^2\ge b_1$. The line element
on $\mathbb{R}^4$ is then written as
\be
\mathrm{d}s^2\left(\mathbb{R}^4\right)
=\sum_{i=1}^4 {\hat x_i^2 \ov r^2-b_i}r^2\mathrm{d}r^2 +
\sum_{i=1}^4 \left(r^2-b_i\right)\mathrm{d}\hat x_i^2\ ,
\label{fbh9}
\ee
where
\be
f=\prod_{i=1}^4 \left(r^2-b_i\right)\ . \ee Also the corresponding harmonic
function in $\mathbb{R}^4$ is \be H^{-1}=N^{-1} f^{1/2}
\sum_{i=1}^4 {\hat x_i^2\ov r^2-b_i}\ .
\label{harm}
\ee
Restricting the discussion to our case, the form of the
distribution density \eqn{dits} dictates that
\be b_1=b_2=0 \ ,
\qq b_3=-\ell_1^2 \ ,\qq b_4=-\ell_2^2\ .
\ee
We found convenient
to parameterize the components of the unit four-vector according to
the decomposition of the $\bf 4$ of $SO(4)$ under doublets and
singlets of $SO(2)\times SO(2)$, i.e. $\bf 4\to (2,1) + (1,2)$, as
\be
\pmatrix{\hat x_1\cr \hat x_2}=\cos\th \pmatrix{\cos\tau\cr
\sin\tau}\ ,\qq \pmatrix{\hat x_3\cr \hat x_4}=\sin\th
\pmatrix{\cos\psi\cr \sin\psi}\ ,
\label{jds8}
\ee
where the
ranges of the angular variables are
\be
0\le \th\le \pi\ ,\qq 0
\le \psi,\ \tau \le 2 \pi \ .
\label{tr2}
\ee
Also for convenience we define the functions
\ba && \D_i
=1+{\ell_i^2\ov r^2}\ ,\qq i=1,2\ ,
\nonumber\\
&& \D = \D_1\D_2 \cos^2\th +\sin^2\th \left(\D_1\sin^2\psi + \D_2 \cos^2\psi\right)\ .
\ea
After some tedious but nevertheless straightforward algebra we find,
from Eqs. \eqn{fbh9} and \eqn{harm}, that the metric in $\mathbb{R}^4$ is
\ba
&& \mathrm{d}s^2\left(\mathbb{R}^4\right)  =  {\D\ov \D_1\D_2} \mathrm{d}r^2 +  r^2 \Big[
\left[\sin^2\th+\cos^2\th \left( \D_1\cos^2\psi+\D_2\sin^2\psi\right)\right]\ \mathrm{d}\th^2
+\cos^2\th \ \mathrm{d}\tau^2
\nonumber\\
&&+\sin^2\th \left(\D_1\sin^2\psi+\D_2\cos^2\psi\right)\ \mathrm{d}\psi^2
+ 2 \sin\th\cos\th\sin\psi \cos\psi(\D_2-\D_1)\ \mathrm{d}\th \mathrm{d}\psi\Big]\ ,
\label{met1}
\ea
whereas for the harmonic function we obtain
\be
H^{-1}=N^{-1} {\D r^2\ov \D_1^{1/2}\D_2^{1/2}}\ .
\label{met2}
\ee
These suffice to compute the ten-dimensional metric from \eqn{jw11}.
The non-zero components of the three-form are found, from \eqn{jw1}, to be
\ba
 H_{r\psi\tau}&=&-2 N {\cos^2\th \sin^2\th\ov \D^2 r} \left[\D_1^2 \sin^2\psi
+\D^2_2\cos^2\psi-\D_1\D_2\left(\D_1 \sin^2\psi+\D_2 \cos^2\psi\right)\right]\ ,
\nonumber\\
H_{r\th\tau}&=& 2 N  {\D_1-\D_2\ov  \D^2 r}\sin\th\cos\th\sin\psi\cos\psi
 \left[\sin^2\th + (\D_1+\D_2-\D_1\D_2)\cos^2\th\right]\ ,
\nonumber\\
H_{\th\psi\tau}& =& N {\cos\th \sin\th\ov \D^2}
\Big(\D_1\D_2(\D_1+\D_2)\cos^2\th
 \nonumber \\
&& + \sin^2\th\left[2 \D_1\D_2 +(2
\D_1\D_2-\D_1-\D_2)\left(\D_1\sin^2\psi+\D_2\cos^2\psi\right)\right]\Big)\
.\ea It will be convenient for later use to construct
an antisymmetric tensor that reproduces this as its field
strength. This can be computed and has non-vanishing
components\footnote{We thank D. Zoakos for providing them.}
\begin{eqnarray} B_{\tau \psi} & =  & N {\D_1 \D_2\cos^2\th\ov \D}\ ,
\nonumber\\
B_{\tau\th} &= & N {\D_1-\D_2\ov \D}
\sin\th\cos\th\sin\psi\cos\psi\ .
\label{ant1}
\end{eqnarray}
As advertised above, this background has, for generic values of these
parameters, a $U(1)\times Z_2$ symmetry, where the former is an
isometry generated by the Killing vector $\del_\tau$. The discrete
factor is generated by the transformation
$\ell_1\leftrightarrow\ell_2$, $\psi\to \pi/2-\psi$ and $\th\to
\pi-\th$. This involves an interchange of the two axes of the ellipsis.

As in \cite{sfet1}, new insight can be obtained by performing a T-duality transformation
with respect to the Killing vector $\del_\tau$. Since we are dealing with
NS--NS fields only, the standard rules of \cite{BUSCHER} suffice.
For the metric we find
\ba
 \mathrm{d}s^2  &=& N\Bigg\{
{\mathrm{d}r^2\ov r^2 \sqrt{\D_1\D_2}} + (\D_1\D_2)^{-1/2}\Big[ \left(\D_1\cos^2\psi
+\D_2\sin^2\psi\right)\ \mathrm{d}\th^2 + {\D\ov \cos^2\th}\ \mathrm{d}\varphi^2
\nonumber\\
&&
 + \D_1\D_2 (\mathrm{d}\psi^2 + 2 \mathrm{d}\varphi \mathrm{d}\psi)
+2 (\D_1-\D_2)\tan\th \sin\psi\cos\psi \ \mathrm{d}\varphi \mathrm{d}\th\Big]\Bigg\}
\nonumber\\
& =& N\Bigg\{{\mathrm{d}r^2\ov r^2 \sqrt{\D_1\D_2}} +
(\D_1\D_2)^{-1/2}\Big[ \left(\D_1\cos^2(\omega-\varphi)
+\D_2\sin^2(\omega-\varphi)\right)\ \mathrm{d}\th^2
\nonumber\\
&&+ \left({\D\ov \cos^2\th}-\D_1\D_2\right)\ \mathrm{d}\varphi^2+ \D_1\D_2 \ \mathrm{d}\om^2
\nonumber\\
&&+ 2 (\D_1-\D_2)\tan\th \sin(\omega-\varphi)\cos(\omega-\varphi) \
\mathrm{d}\varphi \mathrm{d}\th \Bigg\} \ ,
\label{jhw}
\ea
where
\be \varphi= {\tau \ov N}\ ,\qq \om = \psi + {\tau \ov N} \ .
\label{gfge1}
\ee
The antisymmetric tensor turns out to be zero and the dilaton
reads
\be \mathrm{e}^{-2 \Phi}= r^2\cos^2\th\ .
\label{jhwdil}
\ee
Notice
that the dilaton does not depend on the ellipsis parameters $\ell_1$
and $\ell_2$.
Most importantly note that due to the ranges \eqn{tr2} and the
redefinitions \eqn{gfge1}, we have the identifications
\be
\om
\equiv \om + {2\pi \ov N} \ ,\qq \varphi \equiv \varphi + {2\pi\ov N}\ .
\label{hrti}
\ee
Therefore the angular coordinates $\om$ and $\varphi$ have orbifold-type identifications.

\subsection{Limiting cases}

There are several limiting cases that are of particular interest.
First, those in
which the ellipsis degenerates into a circle and a bar,
constructed  in \cite{sfet1} and \cite{basfe2}, respectively and which we review below
for completeness.
In addition, we will examine a new limit in which one of the axes of the ellipsis
becomes very large compared to the other. In that case the ellipsis can be viewed
as two infinitely long bars kept at a fixed distance apart.
In these limiting cases the isometry group gets enlarged.

\subsubsection{Branes on the circle}

When $\ell_1^2=\ell_2^2$ we get the solution for NS5-branes uniformly distributed on a ring,
constructed in \cite{sfet1}.\footnote{Actually this background was constructed before
in \cite{SfeRest} in studies of the interplay between T-duality and supersymmetry, but
the NS5-brane interpretation was given in \cite{sfet1}.}
In this limit, using Eqs. \eqn{met1} and \eqn{ant1} we find that the background becomes
\ba
\mathrm{d}s^{(0)2}
& = & N\left[\mathrm{d}\rho^2+\mathrm{d}\th^2+{1\ov \S}(\tanh^2 \r
\mathrm{d}\tau^2 + \tan^2\th \mathrm{d}\psi^2)\right]\ ,
\nonumber\\
 B^{(0)}_{\tau\psi}& = &{N\ov \S}\ ,
\nonumber\\
 \mathrm{e}^{-2\Phi^{(0)}} & = & \S \cosh^2\r \cos^2\th\ ,
\label{meci}
\ea
where we changed the radial variable as $r=\ell_1\sinh\r$ and
\be
\S=\tanh^2\r \tan^2\th+1\ .
\ee
Note that for later convenience we have included an extra index in the
fields to emphasize that they represent the leading order results
in the expansion of the general ellipsis background
around the circle case with $\ell_1^2=\ell_2^2$.
The isometry group in Eq. \eqn{meci} is
$U(1)\times U(1)$ with Killing vectors
$\del_\tau$ and $\del_\psi$, where the two factors correspond to rotations
in the planes perpendicular as well well on
the plane of the NS5-brane distribution. The
second $U(1)$ is a continuous approximation to the $Z_N$ discrete
symmetry that actually leaves the exact background invariant.
The exact form of the background, without passing to the continuous approximation
we employ here, was computed in \cite{sfet1}.
Finally, from Eq. \eqn{jhw}, its T-dual is
\ba
\mathrm{d}s^{(0)2} &=& N \left[ \mathrm{d}\rho^2+\coth^2\rho \mathrm{d}\om^2 +\mathrm{d}\th^2 +
\tan^2\th \mathrm{d}\varphi^2 \right]\ ,
\nonumber\\
\mathrm{e}^{-2 \Phi^{(0)}} & = &\sinh^2\r \cos^2\th\ ,
\label{hjds2}
\ea
with the identifications \eqn{hrti}. This is the background
corresponding to a $Z_N$ orbifold of the exact CFT
$SU(2)/U(1)\times SL(2,\mathbb{R})/U(1)$.

\subsubsection{Branes on a bar}

When $\ell_1\to 0$, the NS5-branes are distributed on a bar of
length $2\ell_2$ along the $x_4$-axis and centered in the origin.
Their distribution will not be uniform in the bar. Instead, taking
in \eqn{dits} the $\ell_1\to 0$ limit we find
\be
\r({\bf x})={1\ov
\pi \ell} \left(1-{x_4^2\ov \ell^2}\right)^{-1/2} \Th\left(1-{x_4^2\ov
\ell^2}\right) \d(x_1)\d(x_2)\d(x_3)\ ,
\ee
where we have, for
notational convenience, dropped the index from $\ell_2$. In this case
the symmetry of the background becomes $SO(3)$ corresponding to
rotations in the three-dimensional space perpendicular to the bar.
However, this symmetry does not become manifest if we just write
down the background or its dual after setting $\ell_1=0$. The reason
is that we have chosen in \eqn{jds8} to parameterize the four-unit
vector in a way that a possible $U(1)\times U(1)$ symmetry in the
background will be manifest, i.e. according to the decomposition
of the $\bf 4$ of $SO(4)$ under the $SO(2)\times SO(2)$ subgroup
as $\bf 4\to (2,1)+(1,2)$. In the case of a distribution on a bar
the convenient parameterization is that corresponding to the
decomposition of the $\bf 4$ of $SO(4)$ under the $SO(3)$ subgroup
as $\bf 4\to 3+1$. The result is given in \cite{basfe2} and is presented
here for completeness. For this, it is convenient to use the
following basis for the unit vectors $\hat x_i$ that define the
three-sphere:
 \be
\pmatrix{\hat x_1\cr \hat x_2}\ = \ \cos\th
 \sin\om \pmatrix{\cos\chi\cr\sin\chi} \ , \quad \hat x_3 \ =\
 \cos\th \cos\om\ , \quad \hat x_4 \ =\  \sin\th\ .
\label{jwoi1}
 \ee
Also for the constants $b_i$ we choose
\be
b_1=b_2=b_3=0\ ,\qq
b_4=-\ell^2\ .
\label{fdj1}
\ee
The expressions following from
\eqn{jw1}, \eqn{dilr}, \eqn{fbh9} and \eqn{harm} for the
four-dimensional transverse part of the metric, the antisymmetric
tensor and the dilaton fields are given explicitly by
\ba
 \mathrm{d}s^2 & =
& N \left(1+{\ell^2\ov r^2}\right)^{1/2}
 \left[{\mathrm{d}r^2\ov r^2 + \ell^2}
+\mathrm{d}\th^2 +{r^2 \cos^2\th\ov r^2 +\ell^2 \cos^2\th}\left(\mathrm{d}\om^2 +\sin^2\om
\mathrm{d}\chi^2\right) \right]\ ,
\nonumber\\
B_{\om\chi} & = & N \sin\om \left(\th + {r^2 \cos\th \sin\th \ov
r^2 +\ell^2 \cos^2\th}\right)\ ,
\nonumber\\
\mathrm{e}^{2\Phi} & = & {\left(1+{\ell^2\ov r^2}\right)^{1/2}\ov r^2+\ell^2\cos^2\th}\ .
\label{jds11}
\ea

\subsubsection{The Eguchi--Hanson metric}

When one of the axis of the ellipsis becomes very large compared
to the other, say $\ell_2\gg\ell_1$, then the distribution density
\eqn{dits} becomes
\be
\r({\bf x})=\lim_{ L\to \infty} {1\ov 4 L}
\Th(L-|x_4|)\left[\d(x_3-\ell)+\d(x_3+\ell)\right]\d(x_1)\d(x_2)\ ,
\ee
where we have dropped for convenience the index in $\ell_1$. This
distribution corresponds to two parallel, infinitely extended
bars each one carrying half of the branes. The limiting procedure
can be directly performed in the background as it is written in
\eqn{met1}--\eqn{ant1}. Indeed, let $\psi=z/\ell_2$ and then take the
$\ell_2\to \infty$ limit which corresponds to a contraction.
In this limit the harmonic function in \eqn{met2} becomes
\be
H={N\ov\ell_2} {\sqrt{r^2+\ell^2}\ov r^2 +\ell^2 \cos^2\th} \ .
\ee
In order to make sense of the limit we have to absorb into
the overall scale the factor $1/\ell_2$. Using the parameterization
\be
\pmatrix{x_1\cr x_2} = r \cos \th \pmatrix{\cos\tau \cr
\sin\tau} \ , \qq x_3 = \sqrt{r^2 +\ell^2} \sin\th \ ,\qq x_4 =
z\sin\th\ , \ee
we find the useful relations \ba && R_+ + R_- =2
\sqrt{r^2+\ell^2} \ , \qq R_+ - R_- = 2\ell \sin\th \ , \qq
R_+ R_- = r^2 +\ell^2 \cos^2\th\ ,
\nonumber\\
&& R_\pm = \sqrt{x_1^2 + x_2^2 + (x_3\pm \ell)^2}\ .
\ea
The result is a metric of the form
\ba
 \mathrm{d}s^2 = H \left( \mathrm{d}x_4^2 + \mathrm{d}{\bf x}^2 \right)\ ,\qq
 H = {N\ov 2}\left({1\ov R_+}+ {1\ov R_-}\right)\ ,
\label{tdeh}
\ea
where we have defined the three-dimensional vector ${\bf x}=(x_1,x_2,x_3)$
The antisymmetric tensor two-form reads:
\ba
&& B = \mathrm{d}x_4 \wedge {\bf \om}\cdot  \mathrm{d}{\bf x} \ ,
\nonumber\\
&&
{\bf \om} = {N/2\ov x_1^2 + x_2^2}
\left({x_3+\ell\ov R_+} + {x_3-\ell\ov R_-} \right)\ (-x_2,x_1,0)\ ,
\label{bsto}
\ea
where we have also absorbed a factor of $1/\ell_2$ into the overall scale as in the
case of the metric above.
Its T-dual with respect to the
Killing vector $\del/\del x_4$ is nothing but the
purely gravitational Eguchi--Hanson metric \cite{EH} written in its
Gibbons--Hawking \cite{GiHa} two-center form
\be
\mathrm{d}s^2 = H^{-1} ( \mathrm{d}x_4 + {\bf \om}\cdot \mathrm{d}{\bf x})^2
+ H \mathrm{d}{\bf  x}^2\ .
\label{ehha}
\ee
Regularity of the metric at the two {\it nuts} at $({\bf x},x_4)=({\bf 0},\pm \ell)$
requires the identification
$x_4\equiv x_4 + 4\pi N$. The symmetry
of the Eguchi--Hanson metric \cite{EH} is $SU(2) \times U(1)/Z_2$. This is not manifest
in the two-center form of the metric but there are global-symmetry considerations
for that as well as an explicit transformation \cite{prasad}
that transforms the metric into the
form presented initially by Eguchi--Hanson.
The background \eqn{tdeh}--\eqn{bsto} inherits
the same symmetry due to the fact that $\del/\del x_4$ is a translational Killing vector.
We note also that \eqn{ehha} is not the background corresponding to
the metric \eqn{jhw} and the dilaton \eqn{jhwdil} after the contraction limit is taken.
The latter gives a background that is not purely gravitational since there is still a non-trivial
dilaton, as it can be easily seen. It can be obtained from the Eguchi--Hanson metric if
the T-duality is performed with respect to a second $U(1)$ corresponding to rotations in
the ($x_1, x_2$) plane.

\section{The deformation from the circle to the ellipsis}\label{ctoe}

In this section we will consider ellipses that deviate slightly
from circles and similarly for their T-duals. We will be
interested in characterizing the perturbation in terms of the
corresponding CFT. We will explicitly show that the perturbation
can be written as bilinears in parafermions, which are the natural
chiral and anti-chiral objects in the $SU(2)/U(1)$ coset model,
dressed appropriately so that the perturbation is marginal.

\subsection{Expansion around the circle}\label{expcir}

First we expand the general
ellipsis background around the circle limiting case when
$\ell_1^2=\ell_2^2$, Eq. \eqn{meci}.\footnote{We also omit, for notational
convenience, writing explicitly the overall factor $N$.
This will be done consistently for the rest of
this paper.} From Eqs. \eqn{met1} and \eqn{met2} we compute the
leading order correction to the background \eqn{meci} to be for the metric
\ba
\mathrm{d}s^{(1)2}  & = &
{\ell_1^2-\ell_2^2\ov 2 \ell_1^2 \cosh^2\r}\Bigg[ \cos 2\psi \left(\mathrm{d}\th^2
+\tan^2\th {\tanh^4\r \mathrm{d}\tau^2-\mathrm{d}\psi^2\ov \S^2}\right)
\nonumber\\
&&-2 \sin2\psi {\tan\th\ov \S} \mathrm{d}\psi \mathrm{d}\th+
\mathrm{d}\rho^2 +{\tanh^2\r\ov \S^2}\left(\mathrm{d}\tau^2 -\tan^2\th\
\mathrm{d}\psi^2\right)\Bigg]
\ ,
\label{jsd9}
\ea
where, as before, we have changed to a new radial variable as $\r=\ell_1\sinh \r$.
For the antisymmetric tensor, the leading correction is
\ba
 B^{(1)}_{\tau\psi}& = & -{\ell_1^2-\ell_2^2\ov 2 \ell_1^2\cosh^2\r}
(1-\cos2\psi){\tanh^2\r \tan^2\th\ov\S^2} \ ,
\nonumber\\
B^{(1)}_{\tau\th} & = & {\ell_1^2-\ell_2^2\ov 2 \ell_1^2\cosh^2\r} \sin
2\psi{\tanh^2\r\tan\th\ov \S} \ .
\label{jf9}
\ea
Similarly, for the T-dual background the leading order correction to the metric is
\ba
 \mathrm{d}s^{(1)2}  &=& {\ell_1^2-\ell_2^2\ov 2 \ell_1^2 \cosh^2\r} \Bigg[\mathrm{d}\rho^2-\coth^2\r \mathrm{d}\om^2
+\cos 2(\omega-\varphi)\left(\mathrm{d}\th^2-\tan^2\th \mathrm{d}\varphi^2\right)\nonumber\\
 &&+ 2\sin 2(\omega-\varphi) \tan\th \mathrm{d}\varphi \mathrm{d}\th \Bigg]\ .
\ea

We observe that the corrections have terms that are invariant
under shifts of $\psi$. These terms can be absorbed by a simple
reparametrization of $\r$. Indeed, let
\be
\r\to \r -{\ell_1^2-\ell_2^2\ov 4} \tanh\r\ .
\ee
This induces a correction to the background \eqn{meci} which should be combined with
\eqn{jsd9} and \eqn{jf9}.
In this way we find that the correction to the metric \eqn{jsd9} becomes (``tot''
stands for the combined deformation and reparameterization)
\ba
\mathrm{d}s^{(1)2}\big|_{\rm tot}
 &=& {\ell_1^2-\ell_2^2\ov 2 \ell_1^2 \cosh^2\r}\times \nonumber \\
 &&\Bigg[ \cos 2\psi \left(\mathrm{d}\th^2
 +\tan^2\th {\tanh^4\r \mathrm{d}\tau^2-\mathrm{d}\psi^2\ov \S^2}\right)-2 \sin 2\psi
 {\tan\th\ov \S} \mathrm{d}\psi \mathrm{d}\th  \Bigg].
\label{h3p}
\ea
Also the
correction to the antisymmetric tensor \eqn{jf9} reads:
\ba
 B^{(1)}_{\tau\psi}\big|_{\rm tot} & = & {\ell_1^2-\ell_2^2\ov 2 \ell_1^2 \cosh^2\r}
 \cos2\psi{\tanh^2\r \tan^2\th\ov\S^2} \ ,
 \nonumber\\
 B^{(1)}_{\tau\th}\big|_{\rm tot}
 & = & {\ell_1^2-\ell_2^2\ov 2 \ell_1^2 \cosh^2\r} \sin 2\psi{\tanh^2\r\tan\th\ov \S}\ .
 \label{h4p}
\ea
For the T-dual background expanding around the $\ell_1^2=\ell_2^2$
case \eqn{jhw} we obtain
\ba
 \mathrm{d}s^{(1)2}\big|_{\rm tot}
= {\ell_1^2-\ell_2^2\ov 2 \ell_1^2 \cosh^2\r} \Bigg[ \cos
2(\omega-\varphi)\left(\mathrm{d}\th^2-\tan^2\th \mathrm{d}\varphi^2\right)
 + 2\sin 2(\omega-\varphi) \tan\th \mathrm{d}\varphi \mathrm{d}\th \Bigg] \ .
\label{js79}
\ea
Note also that for the variable $r$ this implies $\d r\sim r$, so that the
dilaton is just shifted by a constant.

\boldmath
\subsection{On parafermions and $SL(2,\mathbb{R})$ primaries}
\unboldmath

Coset theories have natural chirally and anti-chirally conserved
objects, which are called parafermions due to their non-trivial
braiding properties. Let's denote them as
 \be
\Psi_\pm^{(1)}\ ,\
 \bar \Psi_\pm^{(1)}  \in {SU(2)_N\ov U(1)}\qq {\rm and} \qq
 \Psi_\pm^{(2)}\ ,\ \bar \Psi_\pm^{(2)}  \in {SL(2,\mathbb{R})_N\ov U(1)}\ .
 \ee
The first pair are also known as compact parafermions and were first introduced
in \cite{paraf}. Their non-compact counterparts are denoted by the second pair
and were found in \cite{lykken}.
They obey the chiral and anti-chiral conservation laws
 \be
 \del_- \Psi^{(i)}_\pm = 0 \ ,\qq \del_+ \bar \Psi^{(i)}_\pm = 0\
 ,\qq i=1,2\
\label{jsd1}
 \ee
and generate infinite dimensional chiral algebras.
Consider next the semiclassical expressions for the chiral
parafermions in terms of space variables \cite{claspara}
 \ba
 \Psi^{(1)}_\pm& = &
 (\del_+\th \mp i \tan\th \del_+\varphi)
\mathrm{e}^{\mp i (\varphi +\phi_1)}\ ,
\nonumber\\
 \Psi^{(2)}_\pm& = &
 (\del_+\r \mp i \coth\r \del_+\om)
\mathrm{e}^{\mp i (\om +\phi_2)}\ .
 \label{ncpar}
 \ea
The phases are
\ba
\phi_1 & = &
-\ha \int^{\s^+}\! J^1_+\mathrm{d}\s^+ + \ha \int^{\s^-} J_-^1 \mathrm{d}\s^-\ ,
\nonumber\\
 \phi_2 & = &
-\ha \int^{\s^-}\! J^2_-\mathrm{d}\s^- + \ha \int^{\s^+}\! J_+^2 \mathrm{d}\s^+\ ,
\label{phi}
\ea
where $J^1_{\pm}$ and $J^2_{\pm}$ are the current components
associated to the Killing vectors $\partial_\varphi$ and
$\partial_\omega$, respectively\footnote{Recall that in general
the currents corresponding to a Killing vector $ \xi$ read
$$
 J_{\pm}^{\xi}=\xi^{\mu} \left(G_{\mu\nu}\mp B_{\mu\nu}\right)\partial_{\pm}x^{\nu}
$$
and obey, on shell, the conservation law
$\del_+ J_{-}^{\xi} + \del_- J_{+}^{\xi}=0$.}
 \be
  J^1_\pm = \tan^2\th \del_\pm \varphi\ ,\qq
 J^2_\pm =
 \coth^2\r \del_\pm \om \ .
\label{jc2}
 \ee
Note that the phases obey on-shell the condition
 \be
\del_+\del_- \phi_i =
 \del_-\del_+ \phi_i \ ,\qq i=1,2\ ,
 \ee
and are well defined, due to the first
of the classical equations of motion
\ba
\del_+ J^1_- + \del_-
J^1_+ = 0 \ ,\qq \del_+\del_- \th - {\sin\th\ov
\cos^3\th}\del_+\varphi \del_-\varphi= 0\ , \ea for the
$SU(2)/U(1)$ coset and \ba \del_+ J^2_- + \del_- J^2_+ = 0 \ ,\qq
\del_+\del_- \r +{\cosh\r\ov \sinh^3\r}\del_+\om \del_-\om = 0 \ ,
\ea
for the $SL(2,\mathbb{R})/U(1)$ coset. The same equations of
motion ensure that for the anti-chiral ones the corresponding
expressions are
 \ba
 \bar \Psi^{(1)}_\pm& = &
 (\del_-\th \pm i \tan\th \del_-\varphi)\
\mathrm{e}^{\pm i (\varphi -\phi_1)}\ ,
\nonumber\\
 \bar\Psi^{(2)}_\pm& = & (\del_-\r \pm i \coth\r \del_-\om)\
\mathrm{e}^{\pm i (\om -\phi_2)}\ .
\label{ncbpar}
\ea
The full set of the classical equations of motion guaranties  also
the conservation laws \eqn{jsd1}. For the readers who are willing to check that, we
note that using current conservation we have that
$\int^{\s^\pm}\!\del_\mp J^i_\pm d\s^\pm = - J^i_\mp$. Therefore,
$\del_\pm\phi_1= \mp J^1_\pm$ and $\del_\pm\phi_2= \pm J^2_\pm$.
Due to the non-local phases attached to them, the classical parafermions are non-local
objects and have non-trivial braiding properties.
The tentative reader will notice that these phases are different from the ones used in the
computation of the Poisson algebra for the classical parafermions (see \cite{claspara}).
In this computation it was important that the parafermions were not dependent on
the past history of the ``time" variable
($-$ for the $\Psi^{(i)}_\pm$ and $+$ for the $\bar\Psi^{(i)}_\pm$).
Hence, for $\Psi^{(i)}_\pm$ the phases were chosen to be twice the terms involving
$J^i_+$ in \eqn{phi}. Similarly, for $\bar\Psi^{(i)}_\pm$ the phases were chosen
to be twice the terms involving $J^i_-$ in \eqn{phi}. This does not affect the
expressions for the derivatives $\del_\pm\phi_{1,2}$ given above and therefore their
conservation properties \eqn{jsd1} remain intact. The choice of phases we have made
facilitates the computations of this paper.

The chiral and anti-chiral energy--momentum tensors for the
$\s$-model corresponding to the background \eqn{hjds2} can be
written as a sum of terms involving the corresponding chiral and
anti-chiral parafermions. In particular, we have that
\be
T_{++}=g_{ij}\del_+x^i\del_+ x^j =
\Psi^{(1)}_+ \Psi^{(1)}_- + \Psi^{(2)}_+ \Psi^{(2)}_-\ ,
\label{eggr}
\ee
obeying $\del_- T_{++}=0$, as it should be and similarly for $T_{--}$.

We are interested in constructing chiral and anti-chiral objects
for the T-dual to \eqn{hjds2} background, namely for the background
\eqn{meci}.
Under the T-duality transformation we have the following map of
worldsheet derivatives
\be
\del_\pm\varphi \to \pm \tilde J^1_\pm\ ,\
\ee
where we have defined
\be
\tilde J^1_\pm = {1\ov \S }\left(\tanh^2\r \del_\pm\tau \mp \del_\pm\psi\right)\ .
\label{jhd2}
\ee
For convenience we also define the currents
\be
\tilde J^2_\pm =
{1\ov \S }\left(\tan^2\th \del_\pm\psi \pm \del_\pm\tau\right)\
\label{jh32}
\ee
and note the transformation of the worldsheet derivatives
\be
\del_\pm \om \to \tanh^2\r \tilde J^2_\pm\ .
\ee
The equations of
motion that follow from varying $\tau$ and $\psi$ are
\be
\del_+\tilde J^i_- + \del_- \tilde J^i_+ = 0 \ ,\qq i=1,2 \ ,
\label{h34}
\ee
whereas those from varying $\th$ and $\r$ are
\ba
&& \del_+\del_-
\th + {\sin\th\ov \cos^3\th} \tilde J^1_+ \tilde J^1_-=0\ ,
\nonumber\\
&& \del_+\del_- \r +{\sinh\r\ov \cosh^3\r}\tilde J^2_+ \tilde
J^2_-=0\ .
\label{h35}
\ea
Using the above, we determine the transformation of the phase factors appearing in
the various expressions for the parafermions, under the
action of T-duality:
\ba
&& \varphi+\phi_1\to -\tau-\tilde \phi\
,\qq \om+\phi_2\to \psi-\tilde \phi\ ,
\nonumber\\
&& \varphi-\phi_1\to \tau+\bar {\tilde \phi}\ ,\qq\phantom{x}
\om-\phi_2\to \psi+\bar{\tilde \phi}\ ,
\ea
where the new phases are
\ba
\tilde \phi & = & -\ha \int^{\s^+}  \left(\tilde J^1_+ + \tilde J^2_+\right)\mathrm{d}\s^+
+ \ha \int^{\s^-}  \left(\tilde J^1_- + \tilde J_-^2\right)\mathrm{d}\s^-\ ,
\nonumber\\
\bar {\tilde \phi} & = & \ha \int^{\s^+} \left(\tilde J^1_+ - \tilde J^2_+\right)\mathrm{d}\s^+ +
\ha \int^{\s^-} \left(-\tilde J^1_- + \tilde J_-^2\right)\mathrm{d}\s^- \ .
\ea
In this way we find that the chiral and the anti-chiral parafermions become
\ba
\tilde \Psi^{(1)}_\pm & = & \left(\del_+\th \mp i \tan\th \tilde J^1_+\right)\
\mathrm{e}^{\pm i (\tau+\tilde \phi)}\ ,
\nonumber\\
\tilde \Psi^{(2)}_\pm & = &  \left(\del_+\r \mp i \tanh\r \tilde J^2_+\right)\
\mathrm{e}^{\mp i (\psi - \tilde \phi)}\ ,
\label{h23}
\ea
and
\ba
\bar {\tilde \Psi}^{(1)}_\pm & = &
\left(\del_-\th \mp i \tan\th \tilde J^1_-\right)\
\mathrm{e}^{\pm i (\tau+\bar {\tilde \phi})}\ ,
\nonumber\\
\bar{\tilde \Psi}^{(2)}_\pm & = &  \left(\del_-\r \pm i \tanh\r \tilde J^2_-\right)\
\mathrm{e}^{\pm i (\psi + \bar {\tilde \phi})}\ ,
\label{h24}
\ea
respectively.
We may check that, on-shell, these parafermions
are indeed chiral and anti-chiral, provided the
equations of motion \eqn{h34} and \eqn{h35}
for the background \eqn{meci} are obeyed.

As an additional
consistency check note that the
energy--momentum
tensor for the $\s$-model corresponding to
the background \eqn{meci} can be written as a sum of terms involving
the corresponding
parafermions. In particular,
\be
\tilde
T_{++}=\tilde g_{ij}\del_+x^i\del_+ x^j =  \tilde\Psi^{(1)}_+
\tilde\Psi^{(1)}_- + \tilde\Psi^{(2)}_+ \tilde\Psi^{(2)}_-\ ,
\ee
as it should be and similarly for $\tilde T_{--}$.

For equal levels $N$ of the $SU(2)$ and $SL(2,\mathbb{R})$ factors in the coset models,
it is well known \cite{paraf,lykken}
that the conformal dimensions of the parafermions are $1\mp 1/N$, the two
different signs corresponding to the compact and non-compact ones, respectively.
It is convenient for this paper to understand them in terms of the energy--momentum
tensor of the theory. Consider first the compact case in
which the energy--momentum tensor is given by the difference of the
Sugawara constructions of the two energy--momentum tensors for the group $SU(2)$ and
its $U(1)$ subgroup. Hence for any operator $\D_{SU(2)/U(1)} = \D_{SU(2)}-\D_{U(1)}$,
where $\D_{U(1)}=m^2/N$ with $m$ the $U(1)$ charge.
The parafermions originate from the $J_\pm$ $SU(2)$ currents
which have $\D_{SU(2)}=1$. Since their charge eigenvalue is $m=\pm 1$, we have
that $\D_{U(1)}=1/N$. Then the result mentioned above follows.
The difference in the sign for the non-compact parafermions is due to the fact that in this
case $\D_{U(1)}=-m^2/N$.

As a final piece of information consider the construction of $SL(2,\mathbb{R})$ WZW theory
affine primaries as composites of group elements.
Let's define the group element $g\in SL(2,\mathbb{R})$ in the spinor representation
as
\ba
g  =   \mathrm{e}^{{i\ov 2} \th_{\mathrm{L}} \s_2 }
\mathrm{e}^{\r\s_1} \mathrm{e}^{{i\ov 2} \th_{\mathrm{R}} \s_2 }\ .
\label{grroP}
\ea
The $SL(2,\mathbb{R})$ algebra generators in terms of Pauli matrices are
$J_0 = \s_2/2$, $J_\pm = \pm (\s_1\mp i \s_3)/2$. Using these, we define
four elements $g_{ab}$, $a,b=\pm$ as
\be
g_{\pm\pm}={\rm Tr}(R_\pm g)\ ,\qq g_{\pm\mp} = \pm {\rm Tr}(J_\pm g)\ ,
\label{ppar1}
\ee
where $R_\pm = \ha (\II\pm \s_2)$.
Explicitly we have that
\be
g_{\pm\pm}= \cosh\r\ \mathrm{e}^{\pm i(\th_{\mathrm{L}}+\th_{\mathrm{R}})/2}\ ,\qq
g_{\pm\mp}= \sinh\r\ \mathrm{e}^{\mp i(\th_{\mathrm{L}}-\th_{\mathrm{R}})/2}\ .
\label{ppar}
\ee
These transform in the $(\ha,\ha)$ representation of
$SL(2,\mathbb{R})_{\mathrm{L}}\times SL(2,\mathbb{R})_{\mathrm{R}}$
with $U(1)$ charges $(\pm \ha ,\pm\ha)$, in all four combinations,
in accordance with their index.
The explicit transformation rules referring to $SL(2,\mathbb{R})_{\mathrm{L}}$ are
\ba
&& \d_0 g_{\pm\pm}=\pm \ha g_{\pm\pm}\ ,\qq \d_0g_{\pm\mp}=\pm \ha g_{\pm\mp} \ ,
\nonumber\\
&& \d_- g_{++}= - g_{-+}\ ,\qq \ \d_+ g_{++}=0\ ,
\nonumber\\
&& \d_- g_{+-}= -g_{--}\ ,\qq \ \d_+ g_{+-}= 0 \ ,
\label{trdf}\\
&& \d_- g_{-+}= 0 \ ,\qq \qq\ \d_+ g_{-+}= g_{++}\ ,
\nonumber\\
&& \d_- g_{--}= 0 \ ,\qq \qq\ \d_+ g_{--}= \d_{+-} \
\nonumber
\ea
and similarly for transformations with respect to $SL(2,\mathbb{R})_{\mathrm{R}}$.
We may construct other irreducible
representations by forming composites of these elements.
In particular, consider the four simple composite objects
\be
A_{++}= {1\ov g_{--}^2}\ ,\qq A_{+-}= {1\ov g_{-+}^2}\ ,\qq A_{-+}= {1\ov g_{+-}^2}\ ,\qq
A_{--}= {1\ov g_{++}^2}\ ,
\label{hfh1}
\ee
which will be very useful as we will soon see.
They clearly have charges $(1,1)$, $(1,-\!1)$, $(-\!1,1)$ and $(-\!1,-\!1)$, respectively.
Using \eqn{trdf} we see that $\d_- A_{++}=\d_- A_{+-} = 0$ and $\d_+ A_{--}=\d_+ A_{-+} = 0$.
Hence, $A_{++}$ forms a lowest weight representation for $SL(2,\mathbb{R})_{\mathrm{L}}$
as well as for $SL(2,\mathbb{R})_{\mathrm{R}}$. The other members of the representation
are obtained by repeatedly acting with $\d_+$ for the left and the right $SL(2,\mathbb{R})$
factor and the charges of these states are accordingly increased.
In terms of $SL(2,\mathbb{R})$ representation theory this state belongs to the positive
discrete series $D^+$ with $j=\bar j=0$ and $U(1)$ charges
$m=\bar m=1$ for both $SL(2,\mathbb{R})$ factors.
Similarly, $A_{+-}$ forms a lowest-weight representation for $SL(2,\mathbb{R})_{\mathrm{L}}$
and a highest-weight representation for $SL(2,\mathbb{R})_{\mathrm{R}}$.
The other members of the representation
are obtained by repeatedly acting with $\d_+$ for the left and $\d_-$
for the right $SL(2,\mathbb{R})$ factor.
In terms of $SL(2,\mathbb{R})$ representation theory this state belongs to the positive
discrete series $D^+$ with $j=0$ and $m=1$ for $SL(2,\mathbb{R})_{\mathrm{L}}$ and
to the negative discrete series $D^-$ with $j=0$ and $\bar m=-1$ for $SL(2,\mathbb{R})_{\mathrm{R}}$.
It is convenient to use for the four states in \eqn{hfh1} a notation that directly relates to
their labelling according to the $SL(2,\mathbb{R})$ representation theory we have mentioned.
In an obvious notation we have\footnote{More general representations were worked
out in a similar fashion in \cite{CLyk} with findings that agree with ours.}
\ba
&& \Phi^{{\rm lw},{\rm lw}}_{0,1,1} = {1\ov g_{--}^2}\ ,\qq
\Phi^{{\rm lw},{\rm hw}}_{0,1,-1} = {1\ov g_{-+}^2}\ ,
\nonumber\\
&& \Phi^{{\rm hw},{\rm lw}}_{0,-1,1} = {1\ov g_{+-}^2}\ ,\qq
\Phi^{{\rm hw},{\rm h w}}_{0,-1,-1} = {1\ov g_{++}^2}\ .
\label{gfhe}
\ea
Since $j=0$ for all of these states, their conformal dimensions, given
by $-j(j+1)/(N-2)$, equal zero.

The background for the $SL(2,\mathbb{R})/U(1)$ follows by considering the vector gauging
of a $U(1)$ subgroup generated by $\s_2$, as it can be seen from
\eqn{grroP}.
In a standard procedure,
the configuration space is reduced by one dimension by a gauge fixing.
Under the vector transformation $\d\th_{\mathrm{L}}=-\e$ and
$\d\th_{\mathrm{R}}= \e$, the $g_{\pm\pm}$ parameters,
as given by \eqn{ppar}, are invariant whereas $g_{\pm\mp}$ are not.
The parafermions are by construction gauge invariant, essentially due to the
dressing of the $J_\pm$ currents with the Wilson lines attached to them.
Because of the gauge invariance we choose a unitary gauge, which amounts
to setting $\th_{\mathrm{L}}=\th_{\mathrm{R}}=\om$. The corresponding background
is given by the relevant part in \eqn{hjds2}. Then
\ba
 g_{\pm\pm}\big|_{\rm g.f.}  =   \cosh\r\ \mathrm{e}^{\pm i\om}\ ,\qq
 g_{\pm\mp}\big|_{\rm g.f.}  =   \sinh\r\
\label{pparge1}
\ea
are the gauged-fixed elements. Since they have $U(1)$ charges $\pm 1$,
the states have conformal dimensions equal to $1/N$.

Under T-duality the factor $\om$ in the states \eqn{gfhe} transforms as
$\om\to \psi -\ha (\tilde \phi -\bar {\tilde \phi})$.

\subsection{The deformation around the circle as a marginal perturbation}
\label{ctoemar}

Using the above formalism, it is easy to show that the correction
\eqn{js79} can be reproduced by adding to the sigma-model action
based on the unperturbed background \eqn{hjds2} the term
 \ba
 \d \cL\vert_{\delta\ell_2}
& = &
 {\ell^2_1-\ell_2^2\ov 4 \, \ell^2_1 }
\left(
\Phi^{{\rm lw},{\rm lw}}_{0,1,1}
\Psi^{(1)}_+ \bar \Psi^{(1)}_-
+ \Phi^{{\rm hw},{\rm hw}}_{0,-1,-1}\Psi^{(1)}_- \bar \Psi^{(1)}_+ \right)
\nonumber\\
& = &
 {\ell^2_1-\ell_2^2\ov 4 \, \ell^2_1 \,
\cosh^2\r }\left(\Psi^{(1)}_+ \bar \Psi^{(1)}_- \mathrm{e}^{2 i\om}
+\Psi^{(1)}_- \bar \Psi^{(1)}_+ \mathrm{e}^{-2 i \om}\right)
\label{delScir}
 \ea
as a perturbation. This is a $(1,1)$ marginal perturbation since
the conformal dimensions add up to one ($1/N+(1-1/N)=1$). There is
however, a delicate point to consider here. In order to preserve
superconformal invariance, the central charge of the theory has to
be $c=6$, out of which two units are due to the free fermions.
Hence, the bosonic part which consists of the cosets we have been
working with has to provide the other four units. This, however, is
only possible if the level of the $SU(2)$ factor is $N$ and that
of the $SL(2,\mathbb{R})$ factor $N+4$ \cite{Kounnas93}. The
supergravity solution is obviously insensitive to this difference
in the levels, since $N$ is assumed to be very large for the
supergravity description to be valid. The above modification
affects the conformal dimension of the operators in \eqn{gfhe}
which becomes $1/(N+4)$ and therefore the total dimension would
receive $1/N^2$ corrections, seizing the perturbation
\eqn{delScir} from being marginal unless we deal with a bosonic
theory. The complete resolution of this issue is postponed for
future work, but we believe that it relies into properly taking
into account the fermionic degrees of freedom. The perturbation
\eqn{delScir} is not factorizable. This is of immense importance
as we know of no other such example in the literature. The r\^ole
of the states in \eqn{gfhe} in dressing the parafermion bilinear
in the perturbation \eqn{delScir} is very important. To appreciate
this point we note that, if the perturbation was purely due to
parafermion bilinears, it would not be marginal and
obviously it would not couple the two factors in the
$SU(2)/U(1)\times SL(2,\mathbb{R})/U(1)$ model. Instead, it was shown
in \cite{FateevInt} that the perturbed $SU(2)/U(1)$
conformal model is integrable and massive, and that it is related to the
$O(3)$ model or it flows in the infrared to the minimal models, depending
on the details of the perturbation.

In the dual background the term (\ref{delScir}) may be
interpreted using the tilded parafermions as
 \ba
 \d \tilde \cL \vert_{\delta\ell_2}
& = &
 {\ell^2_1-\ell_2^2\ov 4 \, \ell^2_1 }
\left(
\tilde \Phi^{{\rm lw},{\rm lw}}_{0,1,1}
{\tilde \Psi}^{(1)}_+ \bar {\tilde \Psi}^{(1)}_-
+ \tilde \Phi^{{\rm hw},{\rm hw}}_{0,-1,-1}{\tilde \Psi}^{(1)}_- \bar {\tilde \Psi}^{(1)}_+
\right)
\nonumber\\
& = &
 {\ell^2_1-\ell_2^2\ov 4\,  \ell^2_1 \, \cosh^2\r }\left(\tilde \Psi^{(1)}_+ \bar
 {\tilde \Psi}^{(1)}_- \mathrm{e}^{2 i\psi -i (\tilde\phi-\bar {\tilde \phi})} +\tilde\Psi^{(1)}_-
 \bar{\tilde\Psi}^{(1)}_+ \mathrm{e}^{-2 i \psi + i (\tilde\phi-\bar {\tilde \phi})}\right)
\ .
\label{TdelScir}
  \ea
These terms precisely reproduce \eqn{h3p} and \eqn{h4p}.
The deformations at hand of the background generated by NS5-branes
distributed over a circle are of a new type, as advertised
previously. In particular, the fact that under the symmetry generated by the Killing
vector $\del_\psi$, the parafermions transform as a doublet
and the fact that \eqn{delScir} is
not a singlet, explains the breaking of the $U(1)$ symmetry
associated with the plane where the
NS5-branes are distributed which occurs as the circle
is deformed into an ellipsis.

It is legitimate to ask at this point why the NS5-branes,
continuously distributed over an ellipsis should give still rise
to an exact theory as for the circle. The main argument in
favor of this is that the background at hand solves the supergravity
equations for any value of $\ell_1/\ell_2$, with the point
$\ell_1 = \ell_2$ being an exact conformal field theory dual to
$SL(2,\mathbb{R})/U(1) \times SU(2)/U(1)$. Furthermore, departure
from $\ell_1 = \ell_2$ is triggered by a conformal operator of
dimension $(1,1)$. One therefore expects that
an underlying exact CFT must exist beyond
$\ell_1 = \ell_2$. Arguing in favor of that is the purpose of next section.

\section{The general perturbation of the ellipsis}

\subsection{Towards generalized parafermions}

We have shown so far that one can move the locus where the
NS5-branes are distributed, from a circle of radius $\ell$ to an
ellipsis with axes $\ell_1$, $\ell_2$. The corresponding metric,
antisymmetric tensor and dilaton background solve, by
construction, the supergravity equations. An important question
is whether this background is perturbatively exact, i.e. if it
receives $\a'\sim 1/N$-corrections. Let's recall that
backgrounds corresponding to two-dimensional $\s$-models with
$(4,4)$ worldsheet generalized extended supersymmetry
\cite{AlvaFre,GHR,PNBW} do not receive $\a'$-corrections
at any order in perturbation theory. This is due to the large extended
supersymmetry which makes
the corresponding counterterms vanish \cite{AlvaFre,HOPAPA}.

The background of the form given in Eq. \eqn{jw11}
has $(4,4)$ extended worldsheet supersymmetry
with the explicit expressions of the complex structures given in \cite{SfeRest}:
\be
F^\pm_i = H \left(\ha\e_{ijk}\ \mathrm{d} x_j \wedge
\mathrm{d}x_k \ \pm\ \mathrm{d}x_i \wedge \mathrm{d}x_4\right)\ , \qq i=1,2,3\ .
\ee
The absence of $1/N$ corrections does not imply the absence of non-perturbative
corrections. In the case of the circle background its exact form, before
the continuum limit for the distribution was taken, contains explicitly such corrections
which, however, are washed out in the continuum limit \cite{sfet1}. These corrections
should be discussed in the spirit of similar studies for the discretized
version of NS5-branes distributed uniformly on an infinite straight line
in \cite{Tong}.
Such non-perturbative corrections are expected to arise for the background
corresponding to the ellipsoidal distribution. Even constructing the
harmonic function beyond the continuum limit \eqn{met2} is however
challenging in this case.

Starting from the circle and going to an almost circular ellipsis
with axes $\ell_1$ and $\ell_2 = \ell_1 + \delta\ell$ is possible
by means of $(1,1)$ marginal operators based on the compact
parafermions of $SU(2)/U(1)$ appropriately dressed with operators
of the non-compact side ($SL(2,\mathbb{R})/U(1)$). Those are
available in the exact conformal field theory of the circle and we
have exhibited them in Sec. \ref{ctoemar}.

A natural question is whether or not we can integrate the above marginal deformation for
finite values of the difference $\ell_2 - \ell_1$. The usual argument for
verifying the integrability of marginal operators \cite{ChauSch} is
valid in the case of left-right factorized ones, namely bilinears
in holomorphic and anti-holomorphic currents. This argument does
not apply, however, in the $(1,1)$ operators appearing in Eqs.
(\ref{delScir}) or (\ref{TdelScir}) which \emph{are not
factorized} because of their non-compact dressing (\ref{gfhe}).
We note that there are many examples of integrated marginal perturbation when the
latter is a current-current perturbation. This was done explicitly for the
background corresponding to the $SU(2)\times SU(2)$ WZW in \cite{HasSen}.

Finding the general conditions under which a non-factorized
$(1,1)$ operator is integrable is a task that goes beyond the
scope of the present work. We will adopt another strategy and
search directly in the background of the ellipsis how to recast a
perturbation $\ell_1\to \ell_1 +\delta\ell_1 $, $\ell_2 \to \ell_2
+ \delta\ell_2$ in the manner (\ref{delScir}) or (\ref{TdelScir}).
As for the circle, the operators driving such a perturbation are
not expected to be factorized and indeed they are not. We will
see, however, that they turn out to exhibit an interesting
\emph{generalized parafermionic structure} valid at any $\ell_1$
and $\ell_2$.

We will concentrate on the background (\ref{jhw}) were the
antisymmetric tensor vanishes and the dilaton does not depend on
the parameters $(\ell_1, \ell_2)$, instead of its T-dual version
\eqn{met1}--\eqn{ant1}. Let us rewrite the metric in a
suggestive form
\begin{eqnarray}
 {\rm ds^2 }
 &=& \left(\frac{{\rm d}r}{r\left(\Delta_1 \Delta_2\right)^{1/4}} + i
 \left(\Delta_1 \Delta_2\right)^{1/4}\, {\rm d}\omega\right)
 \left(\frac{{\rm d}r}{r\left(\Delta_1 \Delta_2\right)^{1/4}} - i
 \left(\Delta_1 \Delta_2\right)^{1/4}\, {\rm d}\omega\right)
  \label{dSl1l2fac}\\
 &+&
  \Bigg(\left(\frac{\Delta_1}{\Delta_2}\right)^{1/4} \left[
\cos\psi\, {\rm d}\theta + \tan \theta \sin \psi \, {\rm d}\varphi
\right]  +i
    \left(\frac{\Delta_2}{\Delta_1}\right)^{1/4}
\left[ \sin\psi\, {\rm d}\theta - \tan \theta \cos \psi \, {\rm
d}\varphi \right]
\Bigg) \nonumber  \\
 & \times & \Bigg(\left(\frac{\Delta_1}{\Delta_2}\right)^{1/4} \left[
\cos\psi\, {\rm d}\theta + \tan \theta \sin \psi \, {\rm d}\varphi
\right] -i\left(\frac{\Delta_2}{\Delta_1}\right)^{1/4} \left[
\sin\psi\, {\rm d}\theta - \tan \theta \cos \psi \, {\rm d}\varphi
\right]
 \Bigg),
\nonumber
 \end{eqnarray}
where $\psi = \omega-\varphi$ as previously -- the coordinates are
$(r, \omega, \varphi, \theta)$.

Possible candidates for the generalized parafermions can be read-off from
expression (\ref{dSl1l2fac}) up to phases and possibly terms that cancel out.
We introduce the following candidates for
\emph{elliptic compact parafermions:}
 \ba
\Psi^{(1)}_\pm &=&
\Bigg[\left(\D_1\ov \D_2\right)^{1/4}(\cos(\omega-\varphi) \del_+\th + \tan\th
\sin(\omega-\varphi) \del_+\varphi)
\nonumber\\
&& \pm i \left(\D_2\ov \D_1\right)^{1/4}(\sin(\omega-\varphi) \del_+\th -
\tan\th \cos(\omega-\varphi) \del_+\varphi)\Bigg]{\rm e}^{\mp i (\om+\phi_1)}
\ ,
\nonumber\\
{\bar \Psi_\pm}^{(1)} &=& \Bigg[\left(\D_1\ov
\D_2\right)^{1/4}(\cos(\omega-\varphi) \del_-\th + \tan\th \sin(\omega-\varphi)
\del_-\varphi)
\nonumber\\
&&  \mp i \left(\D_2\ov \D_1\right)^{1/4}(\sin(\omega-\varphi)
\del_-\th - \tan\th \cos(\omega-\varphi) \del_-\varphi)\Bigg]{\rm
e}^{\pm i (\om -\phi_1)}\ ,
 \label{gerne}
 \ea
as well as their non-compact companions:
 \ba
\Psi^{(2)}_\pm &=& \left[{\del_+ r\ov r (\D_1 \D_2)^{1/4} } \mp i
(\D_1\D_2)^{1/4} \del_+\om \right] \mathrm{e}^{\mp i (\om
+\phi_2)}\ , \nonumber
\\
{\bar \Psi_\pm}^{(2)} & =& \left[{\del_- r\ov r (\D_1 \D_2)^{1/4}
} \pm i (\D_1\D_2)^{1/4} \del_-\om\right] \mathrm{e}^{\pm i (\om -\phi_2)}\ .
\label{bgernenc}
\ea
A noticeable difference with respect to the usual coset
parafermions is that compact and non-compact ones are no longer
decoupled in the ellipsis. This is expected since the metric \eqn{jhw} is not
factorized into two parts.

The phases $\phi_1$ and $\phi_2$ can still be written as in
(\ref{phi}). However, the currents $J_{\pm}^1$ and $J_{\pm}^2$ are
no longer those in
(\ref{jc2}). They must coincide
in the limit $\Delta_1 \to \Delta_2$ where we recover the circle,
in order for the elliptic parafermions
(\ref{gerne})--(\ref{bgernenc}) to match the ordinary parafermions
(\ref{ncpar}) and (\ref{ncbpar}) in this limit:
\begin{eqnarray}
\lim_{\Delta_1 \to \Delta_2}  J_{\pm}^1&=&
g_{\varphi\theta}\vert_{\Delta_1 = \Delta_2}\partial_{\pm}\theta +
 g_{\varphi\varphi}\vert_{\Delta_1 = \Delta_2}\partial_{\pm}\varphi =
 \tan^2\th \del_\pm \varphi\ ,
\nonumber
\\
\lim_{\Delta_1 \to \Delta_2} J_{\pm}^2 &=&
g_{\omega\omega}\vert_{\Delta_1 = \Delta_2} \partial_{\pm}
 \omega=\coth^2 \r \del_\pm \om\ ,
\label{phase12mlim}
\end{eqnarray}
that is we obtain the current components of \eqn{jc2}.
We will come back to the determination of $J_{\pm}^1$ and
$J_{\pm}^2$ when analyzing the chirality properties, in Sec.
\ref{acdef}.

\subsection{The deformation}\label{acdef}

In order to determine the would-be marginal operator at generic
$\ell_i$'s, we must analyze how the action behaves under
deformations of the ellipsis. Using the elliptic parafermions
introduced previously, the Lagrangian of the sigma model in the
background (\ref{dSl1l2fac}) takes the form
\begin{eqnarray}
\mathcal{L}&=&\left(G_{\mu\nu}+B_{\mu\nu}\right)\partial_+x^{\mu}\,
\partial_-x^{\nu}\nonumber \\
&=&\frac{1}{2}\bigg(\Psi^{(1)}_+ \bar \Psi^{(1)}_+ {\rm e}^{2
i\phi_1} +\Psi^{(1)}_- \bar \Psi^{(1)}_- {\rm e}^{-2 i \phi_1}
+\Psi^{(2)}_+ \bar \Psi^{(2)}_+ {\rm e}^{2 i\phi_2} +\Psi^{(2)}_-
\bar \Psi^{(2)}_- {\rm e}^{-2 i \phi_2}\bigg).
\label{genaction}
\end{eqnarray}
Furthermore, the energy--momentum tensor takes again the form \eqn{eggr}.

One can analyze the behavior of $\mathcal{L}$ under a general
deformation of the ellipsis, $\ell_1 \to \ell_1 + \delta\ell_1$
and $\ell_2 \to \ell_2 + \delta\ell_2$. This amounts to
considering $\Delta_1 \to \Delta_1 + \delta \Delta_1$ and
$\Delta_2 \to \Delta_2 + \delta \Delta_2$.
The variation of the action \eqn{genaction} is bilinear in the elliptic parafermions,
as the action itself:
\begin{eqnarray}
\delta \mathcal{L} \vert_{\delta \Delta_1, \delta \Delta_2}
&=&\frac{1}{4}\left(\frac{\delta\Delta_1}{\Delta_1}-\frac{\delta\Delta_2}{\Delta_2}\right)
\left(\Psi^{(1)}_+ \bar \Psi^{(1)}_- {\rm e}^{2 i\omega}
+\Psi^{(1)}_- \bar \Psi^{(1)}_+ {\rm e}^{-2 i \omega}\right)\nonumber \\
&&-\frac{1}{4}\left(\frac{\delta\Delta_1}{\Delta_1}+\frac{\delta\Delta_2}{\Delta_2}\right)
\left(\Psi^{(2)}_+ \bar \Psi^{(2)}_- {\rm e}^{2 i\omega}
+\Psi^{(2)}_- \bar \Psi^{(2)}_+ {\rm e}^{-2 i \omega}\right).
\label{gendelactionddel}
\end{eqnarray}

Similarly to the discussion in Sec. \ref{expcir}, some of this general
deformation turns out to be
equivalent to a coordinate transformation $r \to r + \delta r$.
One easily sees that for rescalings $\delta r = \varepsilon r$,
the variation of the action $\delta \mathcal{L} \vert_{\rm repar}$
can be written in a factorized form, similar to
(\ref{gendelactionddel}). By appropriately tuning $\varepsilon$
versus $\delta\ell_1$ and $\delta\ell_2$, one can possibly
simplify $\delta \mathcal{L} \vert_{\rm tot}= \delta \mathcal{L}
\vert_{\rm repar} + \delta \mathcal{L} \vert_{\delta \Delta_1,
\delta \Delta_2}$. In particular, $\delta \mathcal{L} \vert_{\rm
tot}$ may vanish for $\ell_2\delta\ell_1 = \ell_1\delta\ell_2$,
which corresponds to deformations that do not alter the shape of
the ellipsis (like $\delta\ell_1 = \delta\ell_2$ respects the
circle $\ell_1 = \ell_2$).

We will not further explore the general structure of $\delta
\mathcal{L} \vert_{\rm tot}$.
The reason is that the generalized parafermions lack two basic properties
which seem incompatible. First, since
for a generic ellipsis one breaks one of the $U(1)$'s available for the
circle, only the
combination $\zeta=\partial_\varphi + \partial_\omega$ remains a
Killing vector with
\begin{equation}\label{zetKil}
  J_{\pm}^\zeta= g_{\varphi\theta}\partial_{\pm}\theta +
 g_{\varphi\varphi}\partial_{\pm}\varphi +
g_{\omega\omega}\partial_{\pm}
 \omega,\end{equation}
satisfying the standard conservation equation $\del_+ J_{-}^{\zeta} + \del_- J_{+}^{\zeta}=0$.
Hence it is not clear which
would be the independent phases $\phi_1$ and $\phi_2$ that enter in the definition of the
generalized parafermions.
The second problem is that the
candidates for generalized elliptic parafermions cannot be made chiral
and anti-chiral.
That is neither $\partial_- \Psi^{(1 \mathrm{\,
or\, 2})}_\pm $ nor $\partial_+ \bar \Psi^{(1 \mathrm{\, or\,
2})}_\pm $ vanish.

Any further classical investigation is of little relevance. We
have good reasons to believe that the operators which deform the
ellipsis at a generic point are exactly marginal, not factorizable though.
In order to prove this statement, we must compute their anomalous
dimensions. This is left for future work.

\section{Conclusions}

We would like to summarize our results. Our starting point was
a geometric plus dilaton and antisymmetric background, generated
by NS5-branes spread over an ellipsis. When the ellipsis degenerates
onto a circle, we know that the background is described in terms of an
exact CFT, T-dual to an arbifold of $SL(2,\mathbb{R})/U(1) \times
SU(2)/U(1)$. We have exhibited (both in the direct and in the T-dual)
the $(1,1)$ conformal operator responsible for triggering the
deformation of the circle into an ellipsis. This operator is of
a new type because it cannot be factorized in holomorphic times
anti-holomorphic dimension-one currents.
On the contrary, the marginal operator appears as
a product of holomorphic and anti-holomorphic parafermions, dressed by a
non-left-right-factorized function of the non-compact fields. We have
proven in the purely bosonic case that the dressing allows for precisely
adjusting the total conformal dimensions to $(1,1)$.

We have finally investigated the existence of a marginal generator all
the way in the elliptic deformation. Our analysis is performed at a classical
level and exhibits a natural generalization of the dimension-two operator of
the circle. The latter is based on what we called {\it elliptic} parafermions that we have
introduced and which mix the compact with the non-compact directions.

Although the computation of the anomalous dimension of the advertised
marginal operator remains to be done, its generic structure strongly suggests
that the background generated by NS5-branes distributed over an ellipsis is
the target space of an exact conformal sigma model. The latter includes among
other limiting cases the Eguchi--Hanson geometry. Therefore, besides the relevance
of our approach in the framework of conformal field theory, it also opens
up new possibilities for description of physically relevant string backgrounds.

\vskip .4 in
\centerline{ \bf Acknowledgments}

\no
We would like to thank V. Dotsenko, D. Orlando and
D. Zoakos for scientific discussions.\\
P.M.~P. acknowledges the University of Patras for
kind hospitality as well partial financial support by
the INTAS contract 03-51-6346 and by the EU under the
contracts MEXT-CT-2003-509661, MRTN-CT-2004-005104 and
MRTN-CT-2004-503369.\\
K.~S. acknowledges CERN and the University of Neuch{\^a}tel for hospitality and financial
support during part of this research. In addition partial support is
provided through the European Community's
program ``Constituents, Fundamental Forces and Symmetries of the Universe''
with contract MRTN-CT-2004-005104,
the INTAS contract 03-51-6346 ``Strings, branes and higher-spin gauge fields'',
the Greek Ministry of Education programs $\rm \P Y\Th A\G OPA\S$ with contract 89194 and
the program $\rm E\Pi A N$ with code-number B.545.



\begin{thebibliography}{99}

\bibitem{Callan}
  C.G. Callan, J.A.~Harvey and A.~Strominger,
  Nucl. Phys. {\bf B359} (1991) 611.

\bibitem{stro1}
{A. Strominger, Nucl. Phys. {\bf B343} (1990) 167, Erratum,ibid.
{\bf B353} (1991) 565.}

\bibitem{AFK}
{I. Antoniadis, S. Ferrara and C. Kounnas, Nucl. Phys. {\bf B421}
(1994) 343, \hfill\break {\tt hep-th/9402073}.}

\bibitem{MS1}
{J.M. Maldacena and A. Strominger, JHEP {\bf 12} (1997) 008, {\tt
hep-th/9710014}.
}
\bibitem{sfet1}
  K.~Sfetsos,
  JHEP {\bf 9901} (1999) 015, {\tt hep-th/9811167}.

\bibitem{Sfetsos:1999pq}
  K.~Sfetsos,
  Fortsch.\ Phys.\  {\bf 48} (2000) 199
  {\tt hep-th/9903201}.

\bibitem{Forste:1994wp}
  S.~Forste,
  Phys.\ Lett.\ B {\bf 338} (1994) 36
  \texttt{hep-th/9407198}.

\bibitem{Forste:2003km}
  S.~Forste and D.~Roggenkamp,
  JHEP {\bf 0305} (2003) 071
  \texttt{hep-th/0304234}.


\bibitem{Israel:2003ry}
  D.~Israel, C.~Kounnas and M.P.~Petropoulos,
  JHEP {\bf 0310} (2003) 028,\hfill\break
  \texttt{hep-th/0306053}.

\bibitem{Israel:2003cx}
  D.~Israel,
  JHEP {\bf 0401} (2004) 042
  \texttt{hep-th/0310158}.

\bibitem{Israel:2004vv}
  D.~Israel, C.~Kounnas, D.~Orlando and P.M.~Petropoulos,
  Fortsch. Phys.  {\bf 53} (2005) 73,
  \texttt{hep-th/0405213}.

\bibitem{Petropoulos:2004ir}
  P.~M.~Petropoulos,
 Fortsch.\ Phys.\  {\bf 53} (2005) 970
 \texttt{hep-th/0412328}.

\bibitem{Israel:2004cd}
  D.~Israel, C.~Kounnas, D.~Orlando and P.~M.~Petropoulos,
  Fortsch.\ Phys.\  {\bf 53} (2005) 1030
  \texttt{hep-th/0412220}.

\bibitem{KKPR}
E.~Kiritsis, C.~Kounnas, P.M.~Petropoulos and J.~Rizos,
Nucl. Phys. {\bf B652} (2003) 165, \texttt{hep-th/0204201}.


\bibitem{Giveon}
  A.~Giveon and D.~Kutasov,
  JHEP {\bf 9910} (1999) 034, {\tt hep-th/9909110}.

\bibitem{KKPRproc}
E.~Kiritsis, C.~Kounnas, P.M.~Petropoulos and
J.~Rizos,
\emph{Five-brane configurations, conformal field
theories and the strong-coupling problem},
{\tt hep-th/0312300}.

\bibitem{Israel}
  D.~Israel, C.~Kounnas, A.~Pakman and J.~Troost,
  JHEP {\bf 0406} (2004) 033,\hfill\break {\tt hep-th/040323}.

\bibitem{EH} T.~Eguchi and A.J.~Hanson,
Phys. Lett. {\bf B74} (1978) 249
and
Annals Phys.  {\bf 120} (1979) 82.

\bibitem{BakSfe}
  I.~Bakas and K.~Sfetsos,
  Nucl. Phys. {\bf B573} (2000) 768, {\tt hep-th/9909041}.

\bibitem{BBS}
 I.~Bakas, A.~Brandhuber and K.~Sfetsos,
  Adv. Theor. Math. Phys.  {\bf 3} (1999) 1657, {\tt hep-th/9912132}.

\bibitem{BUSCHER} T.~Buscher,
Phys. Lett. {\bf B194} (1987) 59 and {\bf B201} (1988) 466.

\bibitem{basfe2}
  I.~Bakas and K.~Sfetsos,
  Fortsch. Phys.  {\bf 49} (2001) 419, {\tt hep-th/0012125}.


\bibitem{prasad}
M.K.~Prasad,
Phys. Lett. {\bf B83} (1979) 310.

\bibitem{SfeRest}
  K.~Sfetsos,
  Nucl. Phys. {\bf B463} (1996) 33, {\tt hep-th/9510034}.

\bibitem{GiHa}
G.W.~Gibbons and S.W.~Hawking, Phys. Lett. {\bf B78} (1978) 430.

\bibitem{paraf}
A.B. Zamolodchikov and V.A. Fateev,
  Sov. Phys. JETP {\bf 62} (1985) 215
  [Zh.\ Eksp.\ Teor.\ Fiz.\  {\bf 89} (1985) 380].

\bibitem{lykken}
  J.D.~Lykken,
  Nucl. Phys. {\bf B313} (1989) 473.

\bibitem{claspara}
K.~Bardacki, M.J.~Crescimanno and E.~Rabinovici,
Nucl. Phys. {\bf B344} (1990) 344.

\bibitem{CLyk}
  S.~Chaudhuri and J.D.~Lykken,
  Nucl. Phys. {\bf B396} (1993) 270, {\tt hep-th/9206107}.

\bibitem{Kounnas93}
  C.~Kounnas,
  Phys. Lett. {\bf B321} (1994) 26, {\tt hep-th/9304102}.

\bibitem{FateevInt}
  V.A.~Fateev and A.B.~Zamolodchikov,
  Phys. Lett. {\bf B271} (1991) 91.

\bibitem{AlvaFre}
  L.Alvarez-Gaume and D.Z.~Freedman,
  Commun. Math. Phys. {\bf 80} (1981) 443.

\bibitem{GHR}
S. Gates, C. Hull and M. Rocek, Nucl. Phys. {\bf B248} (1984) 157.

\bibitem{PNBW}
P. van Nieuwenhuizen and B. de Wit, Nucl. Phys. {\bf B312} (1989) 58.

\bibitem{HOPAPA} P. Howe and G. Papadopoulos, Nucl. Phys. {\bf B289} (1987) 264 and
Class. Quant. Grav. {\bf 5} (1988) 1647.

\bibitem{Tong}
  D.~Tong,
  JHEP {\bf 0207} (2002) 013, {\tt hep-th/0204186}.

\bibitem{ChauSch}
S. Chaudhuri and J.A. Schwartz, Phys. Lett. {\bf B219} (1989) {291}.

\bibitem{HasSen}
  S.F.~Hassan and A.~Sen,
  Nucl. Phys. {\bf B405} (1993) 143, {\tt hep-th/9210121}.


\end{thebibliography}
\end{document}